\documentclass[twocolumn]{aastex631}

\usepackage[english]{babel}
\usepackage[utf8]{inputenc}
\usepackage{amsmath}
\usepackage{graphicx}
\usepackage{natbib}
\usepackage{comment}
\usepackage{gensymb}
\usepackage{amssymb}
\newcommand{\GRS}{GRS 1915+105}
\newcommand{\mida}{MID06a}
\newcommand{\midb}{MID06b}
\newcommand{\midc}{MID06c}
\newcommand{\mca}{MCC06a}
\newcommand{\mcb}{MCC06b}
\newcommand{\mcc}{MCC06c}
\newcommand{\diskbb}{\texttt{diskbb}}
\newcommand{\bhspec}{\texttt{bhspec}}
\newcommand{\comptt}{\texttt{comptt}}
\newcommand{\kerrbb}{\texttt{kerrbb}}
\newcommand{\kerrbbtwo}{\texttt{kerrbb2}}
\newcommand{\phabs}{\texttt{phabs}}

\newcommand{\varabs}{\texttt{varabs}}
\newcommand{\smedge}{\texttt{smedge}}
\newcommand{\edge}{\texttt{edge}}
\newcommand{\gaussian}{\texttt{gaussian}}
\newcommand{\nthcomp}{\texttt{nthcomp}}
\newcommand{\pow}{\texttt{powerlaw}}
\newcommand{\simpl}{\texttt{simpl}}
\newcommand{\solarmasses}{$M_\odot$}
\newcommand{\RXTE}{\textit{RXTE}}
\newcommand{\chisquared}{$\chi_{\nu}^2$}
\newcommand{\fcol}{$f_{\rm{col}}$}
\newcommand{\ModelDiskbbNthcomp}{\textbf{\diskbb{}+\nthcomp{}}}
\newcommand{\ModelDiskbbPow}{\textbf{\diskbb{}+\pow{}}}
\newcommand{\ModelPBhspecSimpl}{\textbf{\texttt{mcc:}\bhspec{}+\simpl{}}}
\newcommand{\ModelPBhspecComptt}{\textbf{\texttt{mcc:}\bhspec{}+\comptt{}}}
\newcommand{\ModelPKerrbbSimpl}{\textbf{\texttt{mcc:}\kerrbb{}+\simpl{}}}
\newcommand{\ModelVBhspecSimpl}{\textbf{\texttt{mid:}\bhspec{}+\simpl{}}}
\newcommand{\ModelVBhspecNthcomp}{\textbf{\texttt{mid:}\bhspec{}+\nthcomp{}}}
\newcommand{\middleton}{MID06}
\newcommand{\mcclintock}{MCC06}
\newcommand{\reid}{R14}

\begin{document}
\title{The black hole spin in GRS 1915+105, revisited}

\author[0000-0003-0148-2817]{Brianna S. Mills}
\email{bri@virginia.edu}
\author[0000-0001-7488-4468]{Shane W. Davis}
\affiliation{Department of Astronomy, University of Virginia, 530 McCormick Road Charlottesville, VA 22904, USA}
\author[0000-0002-8183-2970]{Matthew J. Middleton} 
\affiliation{Department of Physics and Astronomy, University of Southampton, Highfield, Southampton SO17 1BJ, UK}

\begin{abstract}
We estimate the black hole spin parameter in GRS 1915+105 using the continuum-fitting method with revised mass and inclination constraints based on the very long baseline interferometric parallax measurement of the distance to this source.  We fit Rossi X-ray Timing Explorer observations selected to be accretion disk-dominated spectral states as described in McClintock et al. (2006) and Middleton et al. (2006), which previously gave discrepant spin estimates with this method.  We find that, using the new system parameters, the spin in both datasets increased, providing a best-fit spin of $a_*=0.86$ for the Middleton et al. data and a poor fit for the McClintock et al. dataset, which becomes pegged at the BHSPEC model limit of $a_*=0.99$.  We explore the impact of the uncertainties in the system parameters, showing that the best-fit spin ranges from $a_*= 0.4$ to 0.99 for the Middleton et al. dataset and allows reasonable fits to the McClintock et al. dataset with near maximal spin for system distances greater than $\sim 10$ kpc. We discuss the uncertainties and implications of these estimates.
\end{abstract}

\section{Introduction}
\label{s:intro}
A soft, apparently thermal emission component is frequently observed in the spectra of black hole candidate X-ray binaries (hereafter BHXRBs) along with a second, harder X-ray component. The soft component is widely believed to be emission from an optically-thick, geometrically thin accretion disk \citep{shakura_sunyaev1973}, while the hard component is thought to be Comptonized emission from hot electrons near the disk (the corona).  Most BHXRBs display variability between spectral states where the relative strengths of these components vary, with the high/soft state referring to cases where the disk component dominates  \citep{remillard_mcclintock2006,doneetal2007}.

When BHXRBs enter strongly disk-dominated high/soft states, one might expect the emission to be well-represented by a bare accretion disk model, which accounts for the relativistic effects on photon emission and the possible change in flow properties at or near the black hole’s innermost stable circular orbit (ISCO). This has motivated a number of relativistic accretion disk spectral models \citep{hanawa1989,gierlinskietal1999,lietal2005,davis_hubeny2006}, which have had success in fitting the spectrum and its variation with accretion rate in the high/soft state of many BHXRBs \citep{gierlinski_done2004,davisetal2005,shafeeetal2006,mcclintocketal2011}.  Placing constraints on the black hole angular momentum, or spin, is a key motivation of many such studies \citep[for a review, see][]{middleton2016}.  The spectrum of the disk is sensitive to the spin through location of the ISCO as well as the relativistic effects on the photon propagation through the black hole spacetime.  This technique of spin measurement is generally referred to as the continuum-fitting method \citep{zhangetal1997,mcclintocketal2011}, which distinguishes it from other spectral-fitting spin measurements such as those that fit the reflected emission features, including the prominent Fe K$\alpha$ line \citep{fabianetal1989}.

The continuum-fitting method has previously been applied to the BHXRB \GRS{}, yielding inconsistent estimates for the black hole spin \citep{mcclintocketal2006,middletonetal2006}.  Although these studies used similar spectral models and fitting methods,  \citet{middletonetal2006} (hereafter \middleton{}) found a more moderate spin parameter ($a_* \sim 0.7$) while \citet{mcclintocketal2006} (hereafter \mcclintock{}) favored high spin ($a_* \gtrsim 0.98$), where we define the dimensionless spin parameter $a_*=J c/(GM^2)$ and $J$ is the angular momentum of the black hole.  This discrepancy in spin can be primarily attributed to differences in the selection of spectra used in their analyses. These differences arise from the difficulties in unambiguously identifying a disk-dominated state in this source, which is famous for its complex variability, with a diversity not generally seen in other low mass X-ray binaries \citep{bellonietal2000} except for the black hole source IGR J17091-3624 \citep{altamiranoetal2011}.

Both studies focused on analysis of spectral observations using data taken by the PCA detector on board the Rossi X-Ray Timing Explorer (hereafter \RXTE{}). \mcclintock{} identified a selection of apparently disk-dominated spectra in the 3-25 keV range based on a number of screening criteria, including RMS variability and hardness ratio, ultimately arriving at 20 candidates.  \middleton{} argued against identifying the \mcclintock{} sample as disk-dominated, instead arguing these observations are more like very high/steep power law states \citep{remillard_mcclintock2006,doneetal2007}, in which a low temperature Comptonization component is present \citep[\middleton{}]{zdziarskietal2005}. Instead, \middleton{} generated a large library of spectra in the 3-20 keV range with 16 second exposures from observations within the $\beta$ and $\kappa$ variability classes of \citet{bellonietal2000}, ultimately identifying 34 disk-dominated candidates.  \mcclintock{}, in turn, criticized this selection process, raising concerns about potential systematic errors arising from the short 16 second exposures while also worrying that the implied luminosity of observations were above or sufficiently close to the Eddington limit as to invalidate the assumptions of the underlying disk model.  In contrast,  \mcclintock{} had focused their spectral analysis on relatively low Eddington observations, where the assumption of a thin accretion disk is more self-consistent.  In fact, the highest luminosity observations among the \mcclintock{} sample showed a trend toward decreasing best-fit spin, in better agreement with the \middleton{} results. The result is that the community has been left to decide for themselves which selection criteria seems preferable, or whether either is robust. Nevertheless, some support for the higher spin estimate of \mcclintock{} is provided by efforts to model the reflection component \citep{milleretal2013}, which also favors nearly maximal spins.

It is notable, however, that both of these previous papers (\mcclintock{}, \middleton{}) made assumptions about the distance and black hole mass that are not well-supported by more recent efforts to constrain these system parameters based on very long baseline interferometric (VLBI) parallax distance measurements \citep[hereafter \reid{}]{reidetal2014}, which place \GRS{} at a smaller distance from us than originally assumed, with a lower black hole mass and a slightly lower inclination. \reid{} report a preliminary analysis of black hole spin $a_*=0.98 \pm 0.01$ (statistical error only), which would be consistent with the previous estimate from \mcclintock{}. Unfortunately, the continuum-fitting analysis is not a primary focus of\reid{} so the results are not described in extensive detail.  Nor does it cover the \middleton{} selected data. Therefore, the goal of this study is to reanalyze both datasets, using these updated system parameters and their associated uncertainties to explore the uncertainty on the best-fitting spin. 

The plan of this work is as follows:  In Section~\ref{s:data} we summarize our methods of data selection and data reduction.  In Section~\ref{s:fitting}, we describe our spectral fitting models and best-fit results.  We discuss our results in Section~\ref{s:discuss}, and summarize our conclusions in Section~\ref{s:conclusions}.


\begin{table*}[t]
\centering
\begin{tabular}{lllll}
Model Name & XSPEC Notation \\
\hline
\ModelDiskbbNthcomp{} & \varabs{}*\smedge{}(\diskbb{}+\nthcomp{}+\gaussian{}) \\
\ModelDiskbbPow{}     & \phabs{}*\smedge{}(\diskbb{}+\pow{}+\gaussian{}) \\ 
\ModelVBhspecSimpl{}  & \varabs{}*\simpl{}(\bhspec{})  \\ 
\ModelVBhspecNthcomp{} & \varabs{}*\smedge{}(\bhspec{}+\nthcomp{}+\gaussian{}) \\
\ModelPBhspecSimpl{}  & \phabs{}*\edge{}*\smedge{}*\simpl{}(\bhspec{}+\gaussian{}) \\ 
\ModelPBhspecComptt{}  & \phabs{}*\edge{}*\smedge{}(\bhspec{}+\comptt{}+\gaussian{}) \\ 
\ModelPKerrbbSimpl{}  & \phabs{}*\edge{}*\smedge{}*\simpl{}(\kerrbb{}+\gaussian{}) \\

\end{tabular}
\caption{List of models used in this paper and their full corresponding XSPEC notation. The prefix ``mid'' refers to models used to fit the selected \middleton{} observations following the same parameter and abundance prescriptions in \middleton{}, and the prefix ``mcc'' is similarly used for the selected \mcclintock{} observations for their parameter and abundance prescriptions (see Section \ref{ss: nonrelativistic}).}
\end{table*}
\label{tab:models}

\section{Data Selection and Reduction}
\label{s:data}

The spectral states of \GRS{} are known to vary quite rapidly on timescales of seconds to days, making it difficult to obtain disk-dominated spectra for continuum-fitting analyses \citep{greineretal1996}. \middleton{} and \mcclintock{} sifted through archival \RXTE{} data and generated large libraries of spectra determined to be disk-dominated. We briefly summarize the key differences in the two data selections, but refer the reader to the respective papers for further details.

\middleton{} generated a large library of spectra in the 3-20 keV range. They then selected out intervals of 16 seconds in which the disk contribution was more than 85\% of the total bolometric luminosity. The 16 second exposures were set by the shortest timing resolution of \RXTE{} and were chosen to avoid the variability seen in the longer exposures. However, \middleton{} note that the variability of \GRS{} can be seen on timescales shorter than 16 seconds and thus require that the 16 second intervals have a rms variability less than 5\%. Ultimately, 34 disk-dominated spectra across 6 \RXTE{} observations in 16 second intervals were identified. These observations are within the $\beta$ and $\kappa$ variability classes of \citet{bellonietal2000}, in which the transition between spectral states is slow. Three of the 34 spectra were chosen by \middleton{} for their continuum-fitting analysis, which we also adopt in this paper: \RXTE{} observation IDs 20402-01-45-03, 10408-01-10-00, and 10408-01-38-00, hereafter referred to as \mida{}, \midb{}, and \midc{}, respectively. The following start and stop times for each observation's 16 second interval used in our data reduction are: \mida{} (116417059 - 116417075), \midb{} (75756947 - 75756963), and \midc{} (87295987 - 87296003) (see \middleton{} their Figure 2). These start and stop times are in \RXTE{} mission elapsed time (seconds). In their analysis, \middleton{} did not include the conventional 1\% systematic error that is often added while performing spectral fitting to account for residuals that can be as large as 1\% in the power law fits to the Crab Nebula. Since these observations are very short, the systematics are not expected to dominate the observations. We found that our results did not change significantly when we included a 1\% systematic error, but nevertheless decided to retain it in the rest of our analysis.

\mcclintock{} generated a large library of observations determined to be in a disk-dominated state using the following criteria: the disk contribution was more than 75\% of the bolometric luminosity (in the 2-20 keV range) with QPOs either absent or weak and only allowing a small rms variability ($<$7.5\%) \cite{mcclintock_remillard2006}. This resulted in 20 candidate disk-dominated observations. In contrast to the three 16 second interval spectra used in the \middleton{} analysis, the spectra used for the \mcclintock{} analysis remained thousands of seconds long. \mcclintock{} identified five of the 20 observations as ``key low-luminosity'' spectra critical to their continuum-fitting analysis, where $L/L_{\mathrm{Edd}} < 0.3$, and $L_{\mathrm{Edd}}$ is the Eddington luminosity. We note that this inferred luminosity and mass depends on the distance to \GRS{}, which given the closer VLBI distance and smaller mass should push the inferred $L/L_{\mathrm{Edd}}$ up. This implies that the \mcclintock{} Eddington ratio criterion was more strict than was required. Contrary to the \mcclintock{} observations, \middleton{} did not impose an Eddington ratio cut-off for their spectra. From the five, key, low-luminosity \mcclintock{} observations, we chose three for our re-analysis: \RXTE{} ObsIDs 10408-01-20-00, 10408-01-20-01, and 30703-01-13-00, hereafter referred to as \mca{}, \mcb{}, and \mcc{}, respectively. Note that the spectral energy range of interest for these spectra is 3-25 keV, which is slightly larger than the 3-20 keV range used for the \middleton{} spectra since the \mcclintock{} observations are much longer and afford more signal-to-noise in the highest energy bins.

We emphasize that neither of the above selection criteria rely on any assumptions about the \GRS{} system parameters. Therefore, we do believe it is necessary to repeat the selection analysis in response to the new VLBI distance constraints. The only exception is that \mcclintock{} chose to focus on a low-luminosity subset of their selected data (with $L/L_{\rm Edd} < 0.3$) for their discussion and we retain that focus here.

We used data reduction software tools from HEASOFT version 6.26.1. Following the same reduction steps in both \middleton{} and \mcclintock{}, Standard-2 PCA spectra were extracted using FTOOLS \texttt{saextrct} and corrected for background using \texttt{runpcabackest}, where all individual xenon gas layers were included. PCU gain variations were not corrected for and xenon layer spectra were not expanded to 256 channels. All spectra were corrected for dead-time. A 1\% systematic error was added to all spectra using \texttt{grppha}. During the data reduction, a Good Time Interval (GTI) is usually specified to screen out undesirable data from events such as earth occultations, passage through the South Atlantic Anomaly, the target being at the edge of the field of view, etc. For the \middleton{} observations, we did not use any GTI criteria as these were only 16 second exposures. For the \mcclintock{} observations, the GTI criterion specified was only data intervals in which all five PCUs were active during the observation.

\section{Spectral Fitting}
\label{s:fitting}

The primary focus of the continuum-fitting method is to apply relativistic accretion disk models such as \kerrbb{} \citep{lietal2005} and \bhspec{} \citep{davisetal2005, davis_hubeny2006} to disk-dominated X-ray spectra and fit for the spin of the black hole. These models can fit for all parameters but degeneracies in how the model parameters affect the spectrum mean that prior knowledge of the distance to the source, $D$, the mass of the black hole, $M$, and the inclination of the accretion disk, $i$, are required for robust constraints. The most recent constraints on these values for the \GRS{} system come from \reid{}: $D = 8.6$ kpc, $M = 12.4$ \solarmasses{}, and $i = 60$\degree, hereafter referred to as the \reid{} preferred values. We utilize XSPEC \citep{arnaud1996} for all of our spectral fitting, and the models used in this paper are collected in Table \ref{tab:models} with their corresponding XSPEC notations.

\subsection{Non-relativistic accretion disk model}
\label{ss: nonrelativistic}
We first confirmed that our data are consistent with those reported in \mcclintock{} and \middleton{} by comparing our fits with the non-relativistic accretion disk model, \diskbb{} \citep{mitsudaetal1984}, with corresponding fits in the two papers. Following the same fit procedure as \middleton{}, we fit \mida{}, \midb{}, and \midc{} tied together with the model \ModelDiskbbNthcomp{} (see Table \ref{tab:models}). This model includes the variable abundance photoelectric absorption model \varabs{}, the \diskbb{} model, the thermal Comptonization model \nthcomp{} \citep{zdziarskietal1996,zyckietal1999}, the smeared edge component \smedge{} \citep{ebisawa1991}, and Gaussian line component \gaussian{}. \middleton{} used abundances from \cite{anders_ebihara1982}, fixing all column densities in \varabs{} to $4.7 \times 10^{22}$ $\mathrm{cm^{-2}}$, except for Si and Fe which were fixed to $16.4 \times 10^{22}$ $\mathrm{cm^{-2}}$ and $10.9 \times 10^{22}$ $\mathrm{cm^{-2}}$, respectively \citep{leeetal2002}. The smeared edge energy was fixed to lie between $6.9-9.0$ keV, following \mcclintock{} (as \middleton{} did not specify any restriction for this parameter), and width fixed at 7.0 keV \cite{shafeeetal2006}. The \gaussian{} line energy was fixed to lie between $6-7$ keV, and the width was fixed at 0.5 keV. We obtained a fit with $\chi^2$ per degree of freedom $=109.66/113$ for all three observations tied together, with \diskbb{} seed photon temperatures of $1.38^{+0.06}_{-0.06}$ keV, $1.68^{+0.06}_{-0.11}$ keV, and $1.93^{+0.14}_{-0.17}$ keV for \mida{}, \midb{}, and \midc{}, respectively, which are within 10\% of the values reported for the same model fit in \middleton{}.

We also fit the \middleton{} observations with \simpl{} \citep{steineretal2009} in place of \nthcomp{}. The \simpl{} model relies on an approximate treatment of inverse Compton scattering, but it assumes the observed soft model component (in this case \diskbb{}) provides the seed photon distribution that is Comptonized to give the hard X-ray emission. With \simpl{}, the additional \smedge{} and \gaussian{} components that are necessary for fitting with \nthcomp{} no longer significantly improve the \diskbb{}+\simpl{} fits. Our best fit \chisquared{} $=173.86/113$, is notably worse with \simpl{} than \nthcomp{}.

Following the fit procedure in \mcclintock{}, the observations \mca{}, \mcb{}, and \mcc{} were all fit separately using the model \ModelDiskbbPow{}. While \middleton{} used \varabs{} for the absorption component, \mcclintock{} used the photoelectric absorption model \phabs{} with relative abundances from \cite{anders_grevesse1989} and a lower fixed column density of $4.0 \times 10^{22}$ $\rm cm^{-2}$ (see Section \ref{ss: absorption} for a discussion on the impact that chosen absorption models and column densities have on our analysis). The smeared edge energy was again fixed to lie between $6.9-9.0$ keV and the width fixed at 7.0 keV. The Gaussian energy was fixed to lie between $6.3-7.5$ keV and the normalization was allowed to go to negative values to allow for absorption, following \mcclintock{}. The \diskbb{} temperatures and best-fit $\chi^2$ per degree of freedom we obtained for each observation are $2.05^{+0.03}_{-0.03}$ keV with \chisquared{} $=48.84/44$ for \mca{}, $2.06^{+0.03}_{-0.03}$ keV with \chisquared{} $=45.76/44$ for \mcb{}, and $2.11^{+0.02}_{-0.02}$ keV with \chisquared{} $=58.08/44$ for \mcc{}. Note that the 44 degrees of freedom reflect that each observation was fit independently of the others, in order to compare with the results from \mcclintock{}. We find that the results are consistent with the fits reported in \mcclintock{}.

\subsection{Relativistic accretion disk models}
\label{ss: relativistic}
We fit both the \mcclintock{} and \middleton{} datasets using the relativistic accretion disk model \bhspec{} \citep{davis_hubeny2006} for $\alpha=0.01$. In their analysis, \middleton{} used the model \ModelVBhspecNthcomp{} with the same \varabs{} prescription and parameter ranges discussed previously in Section \ref{ss: nonrelativistic}.  Instead, we chose to fit \mida{}, \midb{}, and \midc{} simultaneously with the model \ModelVBhspecSimpl{}. This model differs from the model \middleton{} used in that we chose to use \simpl{} \citep{steineretal2009} to fit the hard X-ray emission rather than \nthcomp{}. We performed fits with \bhspec{} and \nthcomp{}, but only report the \simpl{} results here since we believe that tying the seed photon distribution to the soft model component is more self-consistent with a physically motivated accretion disk model.  Furthermore, when used with \bhspec{}, we do not consistently find best-fit results that are disk-dominated since \nthcomp{} can account for a fraction of the softer emission when the temperature of the Comptonizing gas $T_e$ is only slightly larger than seed photon temperature. \simpl{} only has two free parameters (photon power-law index and photon scattered fraction) and we also drop the additional \smedge{} and \gaussian{} components, which do not significantly improve the fit when \simpl{} is used. We constrained the \simpl{} photon index to $\Gamma > 2$. If $\Gamma < 2$ is allowed, fits with high \simpl{} scattering fractions are favored for observation \midc{}, resulting in almost all of the Comptonized emission being present outside the limit of our data $E > 20$ keV. Hence, instead of fitting a power law in the hardest observed X-ray channels, \simpl{} simply depresses the \bhspec{} model flux fitting the softer photons, effectively renormalizing it. Fixing the values for the mass, distance, and inclination in \bhspec{} to the values \middleton{} assumed ($M = 14$ \solarmasses{}, $D =$ 12.5 kpc, and $i = 66$\degree), we find that \midc{} became pegged at the luminosity limit of \bhspec{} ($L/L_{\rm{Edd}} = 1.77$), causing the spin to unrealistically drop to 0. Fixing the mass, distance, and inclination to the new \reid{} preferred values, we obtained a reasonable fit with \chisquared{} $=136.3/125$ and a moderately high spin of $a_* = 0.863^{+0.014}_{-0.015}$. The best-fit parameters are reported in Table \ref{tab:bestfit_middleton}, and the three spectra fit simultaneously with the model \ModelVBhspecSimpl{} assuming the \reid{} preferred values are shown in Figure \ref{fig:middleton_three}.

When we fit only observation \mida{}, the closest in luminosity to the \mcclintock{} observations, the best-fit preferred a high \simpl{} scattering fraction and a significantly lower spin. We found that when separately fitting \mida{}, \midb{}, and \midc{} assuming the \reid{} preferred values and the model \ModelVBhspecSimpl{}, the two lower luminosity datasets (\mida{} and \midb{}) preferred higher scattering fractions with lower spins, whereas the higher luminosity dataset (\midc{}) preferred a lower scattering fraction with a slightly higher spin compared to the best-fit values when all three datasets were tied together.  This preference for high scattering fraction in the lower luminosity observations is partially attributable to \simpl{} depressing the flux of the \bhspec{} model at soft X-ray energies, an effect that is absent when additive models like \nthcomp{} or \comptt{} are fit for the harder X-ray component.

\begin{table*}[tbh!]
\centering
\begin{tabular}{lllll}
Model Component & Parameter & \mida{} & \midb{} & \midc{} \\
\hline
\simpl{}    & $\Gamma$              & $3.52_{-0.33}^{+0.34}$    & $4.15_{-0.48}^{+0.47}$    & $5.00^\tablenotemark{a}$  \\
            & $f_{\rm{sc}}$         & $0.22_{-0.07}^{+0.05}$    & $0.33_{-0.10}^{+0.12}$    & $0.24_{-0.01}^{+0.06}$  \\
\bhspec{}   & $L/L_{\mathrm{Edd}}$  & $0.30_{-0.01}^{+0.01}$    & $0.47_{-0.01}^{+0.01}$    & $0.87_{-0.01}^{+0.01}$  \\
            & $a_*$                 & $0.863_{-0.015}^{+0.014}$ &     (tied)       & (tied) \\
            &\chisquared{} & 136.3/125
\end{tabular}

\tablenotetext{a}{Parameter was completely unconstrained.}

\caption{Best-fit values for the three \middleton{} observations, \mida{}, \midb{}, and \midc{} tied together and fit with the model \ModelVBhspecSimpl{} (see Table \ref{tab:models}). From top to bottom, the parameter values are the \simpl{} model photon index, fraction of scattered photons, the \bhspec{} luminosity, dimensionless spin parameter, and the reported $\chi^2$ per degree of freedom for the entire fit. The mass, distance, and inclination were fixed at the \reid{} preferred values (12.4 \solarmasses{}, 8.6 kpc, and 60\degree).}
\end{table*}\label{tab:bestfit_middleton}

\begin{figure}
    \centering
    \includegraphics[width=\columnwidth]{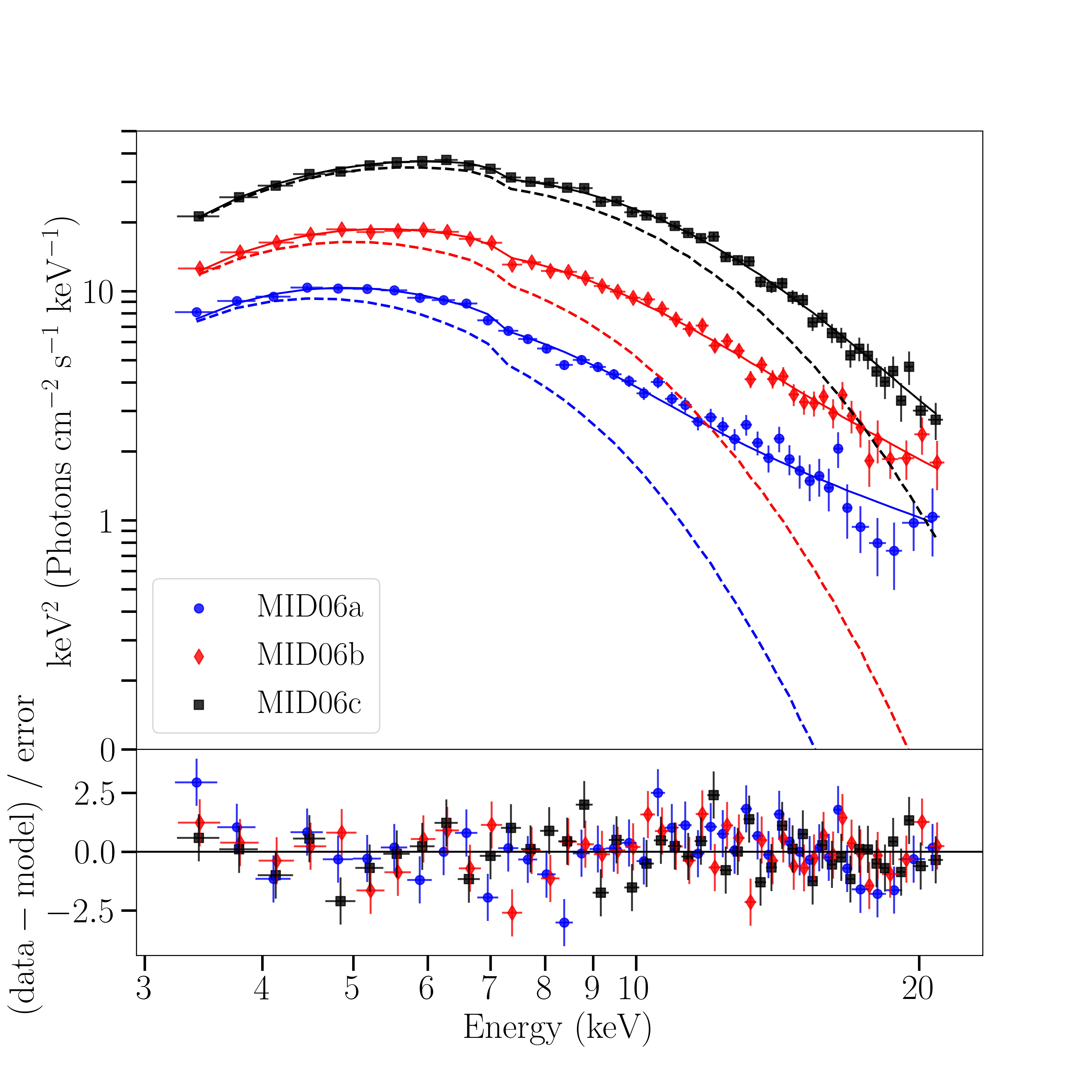}
    \caption{The top panel shows the three \middleton{} \RXTE{} observations, \mida{} (blue circles), \midb{} (red diamonds), and \midc{} (black squares), tied together and fit with the model \ModelVBhspecSimpl{}. The mass, distance, and inclination in \bhspec{} were fixed to the \reid{} preferred values for the \GRS{} system ($M = 12.4$ \solarmasses{}, $D = 8.6$ kpc, and $i = 60$\degree). The solid lines show the total model fit for each observation, and the dashed lines show the disk contribution for each spectral fit by setting the \simpl{} scattering fraction to zero. The bottom panel shows the respective fit residuals.} 
\end{figure}\label{fig:middleton_three}

In their analysis, \mcclintock{} used a variation of \bhspec{} and \kerrbb{} which they call \kerrbbtwo{} to fit the soft X-ray continuum. This hybrid model includes the returning radiation and limb darkening capabilities from \kerrbb{} while constraining the color correction factor, or spectral hardening factor, \fcol{} using look-up tables generated from \bhspec{}. They also used the Comptonization model \comptt{} \citep{titarchuk1994} to fit the hard X-ray spectral component. To stay consistent in our re-analysis of the two papers, we chose to use \simpl{} for the \mcclintock{} data as we did for the \middleton{} data. We first fit \mca{}, \mcb{}, and \mcc{} tied together using only \phabs{}, \bhspec{}, and \simpl{}, fixing the mass, distance, and inclination to the values previously assumed by \mcclintock{} ($M = 14$ \solarmasses{}, $D = 11$ kpc, and $i = 66$\degree). We were unable to obtain a reasonable fit, with \chisquared{} $= 4090.6/150$. We then added the \smedge{} and \gaussian{} components, along with an addition absorption edge component, \edge{}, which \mcclintock{} used to improve their fit results. Here we adopt the same \edge{} prescription, fixing the energy to lie between $8.0-13.0$ keV. We also kept the same abundance and parameter constraints as outlined in Section \ref{ss: nonrelativistic}, and label this model \ModelPBhspecSimpl{}. Adding these components significantly improved the fit results with \chisquared{} $=139.4/132$ and provided a similar spin ($a_* \sim 0.977$) when compared to the spin reported by \mcclintock{} ($a_* \gtrsim 0.98$). Keeping these model components in our fit, we then fixed the mass, distance, and inclination to the \reid{} preferred values, but were unable to get a good fit with \chisquared{} $=835.5/132$ and the $a_*$ became pegged at the maximum spin allowed by \bhspec{} ($a_* = 0.99$). Note that the 132 degrees of freedom reflects all three observations fit simultaneously with the spin parameter tied across all observations.

In contrast to \bhspec{}, the relativistic accretion disk model \kerrbb{} allows the color correction factor \fcol{} to vary as a free parameter. However, fitting for $a_*$ while allowing \fcol{} to be free did not give reliable spin estimates, as the two parameters share a strong degeneracy \citep{salvesen_miller2020}. We discuss fitting \kerrbb{} to the selected \mcclintock{} observations fixing \fcol{} at different values in the next section.

\subsection{Color correction factor}
\label{ss:colorcorrection}

\begin{figure}
\centering
\includegraphics[width=\columnwidth]{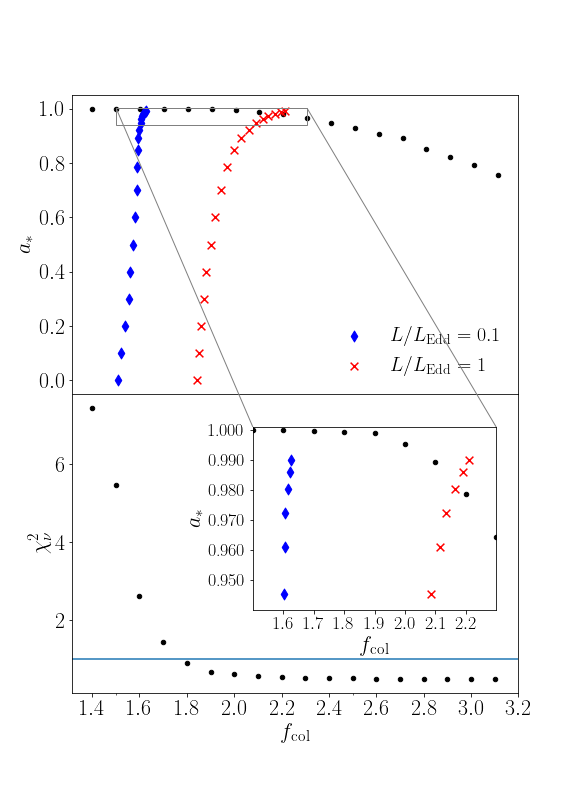}
\caption{Plot showing the range of color correction factor values \fcol{} for the three chosen \mcclintock{} observations \mca{}, \mcb{}, \mcc{} tied together and fit with the \ModelPKerrbbSimpl{} model, shown as black dots. The corresponding best-fit dimensionless spin parameter $a_*$ is shown in the top panel, and \chisquared{} (126 degrees of freedom) is shown in the bottom panel. These fits were calculated assuming a mass, distance, and inclination fixed at the \reid{} preferred values ($M = 12.4$ \solarmasses{}, $D = 8.6$ kpc, and $i = 60$\degree). Estimates for \fcol{}, found by fitting \kerrbb{} to \bhspec{} are overplotted for a fixed $L/L_{\mathrm{Edd}} = 0.1$ (blue diamonds) and a fixed $L/L_{\mathrm{Edd}} = 1$ (red X's). The horizontal line in the bottom panel shows \chisquared{} $=1$ for reference.} \label{fig:color_correction}
\end{figure}

We fit \mca{}, \mcb{}, and \mcc{} tied together and fit with the model \ModelPKerrbbSimpl{}. The same parameter restrictions for the \edge{}, \smedge{}, and \gaussian{} components discussed in Section \ref{ss: relativistic} were again used in this model. For the \kerrbb{} parameters, we assumed zero torque at the inner boundary, limb-darkening, and self-irradiation. The mass, distance, and inclination were fixed at the \reid{} preferred values. We then fixed \fcol{} at different values ranging between $1.4-3.1$ and show each resulting best-fit $a_*$ and \chisquared{} (126 degrees of freedom) plotted as black dots in Figure \ref{fig:color_correction}. Over-plotted in the figure are two estimates of \fcol{} obtained by fitting \kerrbb{} to \bhspec{}: one for fixed $L/L_{\mathrm{Edd}} = 0.1$ (shown as blue diamonds), and one for fixed $L/L_{\mathrm{Edd}} = 1$ (shown as red X's). We obtain the \fcol{} estimates by running the XSPEC fakeit command to generate artificial datasets with a \phabs{}*\bhspec{} model, the response from the MCC06b observation, and assuming an exposure of $10^4$ seconds. We then fit these datasets with \phabs{}*\kerrbb{}, fixing $a_*$, $M$, $D$, $i$, $n_{\rm{H}}$ and the accretion rate to match the values assumed in the faked spectrum, but allowing \fcol{} to be a free parameter.

\begin{figure}
    \centering
    \includegraphics[width=\columnwidth]{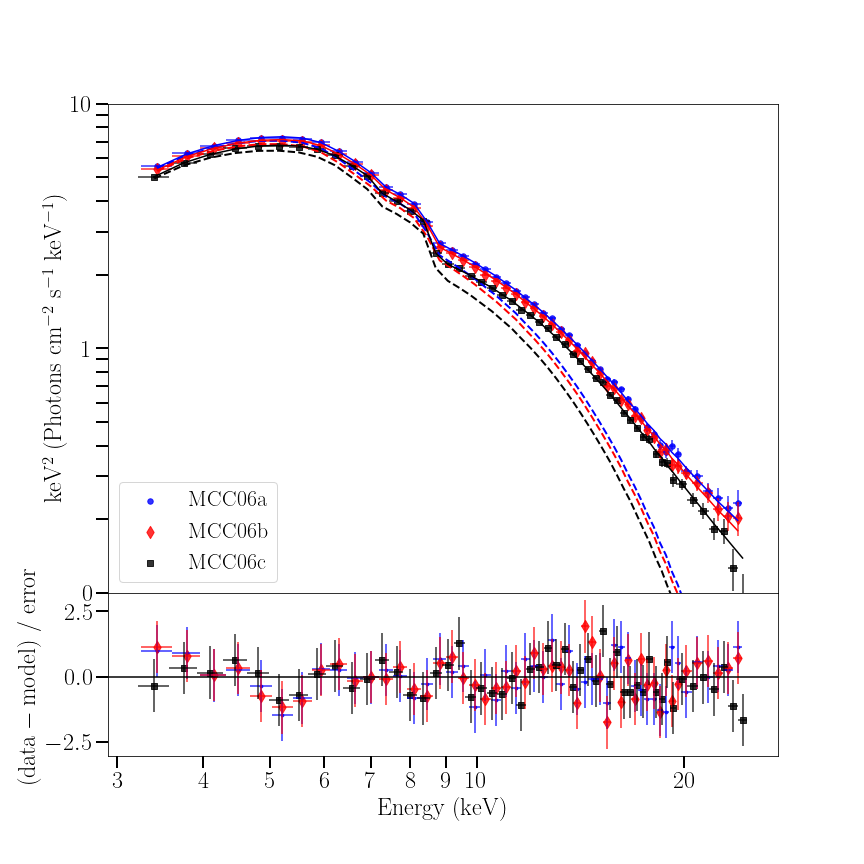}
    \caption{The top panel shows the three \mcclintock{} \RXTE{} observations \mca{}, \mcb{}, \mcc{} tied together and fit with the model \ModelPKerrbbSimpl{}. The mass, distance, and inclination were fixed at the \reid{} preferred values for \GRS{} ($M = 12.4$ \solarmasses{}, $D = 8.6$ kpc, and $i = 60$\degree). The \kerrbb{} color correction factor, or spectral hardening factor, was arbitrarily fixed at \fcol{} $=2.0$. The dashed lines show the contribution of the disk emission for each spectral fit by setting the \simpl{} scattering fraction to zero. The bottom panel shows the fit residuals for each spectrum.} \label{fig:mcclintock_three}
\end{figure}

A reasonable fit can be obtained for color correction factors \fcol{} $\gtrsim 1.7$ if $a_* = 0.999$. Lower spins only provide acceptable fits with higher values of \fcol{}, but sufficiently high values of \fcol{} only occur for $L/L_{\rm{Edd}}$ greater than inferred for the \mcclintock{} data. A representative best-fit to the three spectra by arbitrarily fixing \fcol{} $=2.0$ is shown in Figure \ref{fig:mcclintock_three} where the best-fit spin is $a_* = 0.995^{+0.002}_{-0.003}$ and \chisquared{}$ =77.6/126$.

\subsection{Exploring System Uncertainties}

The uncertainty on the best-fit spin depends directly on the uncertainties in the distance, mass, and inclination. We explored this uncertainty in parameter space by fixing the mass, distance, and inclination at a range of different values above and below the \reid{} preferred values. A distance was randomly sampled from a Gaussian distribution centered on 8.6 kpc, with a 2.0 kpc width chosen to approximately match their uncertainty. From \reid{}, the  dependence of the inclination on a given distance is constrained from VLBI proper motion constraints, assuming ballistic trajectories for the plasma emitting in the jet.  This gives
\begin{equation}
    \tan{i}= \left( \frac{2D}{c}\right) \left(\frac{\mu_\mathrm{a}\mu_\mathrm{r}}{\mu_\mathrm{a} -\mu_\mathrm{r}}\right),\label{eq:inclination}
\end{equation}
where $i$ is the inclination of the accretion disk with respect to the line of sight ($i=0$ is a face-on disk), $D$ is the distance to the black hole, $\mu_a$ is the apparent speed of the approaching radio jet, and $\mu_r$ is the apparent speed of the receding radio jet. The values for $\mu_a$ and $\mu_r$ were also sampled from Gaussian distributions centered on their reported values of 23.6 $\pm$ 0.5 milliarcseconds/yr and 10.0 $\pm$ 0.5 milliarcseconds/yr, respectively (\reid{}). The mass of the black hole is then determined by using the inclination from equation (\ref{eq:inclination}) in the following expression:

\begin{equation}
    M = \frac{\mathcal{M}}{\sin^3{i}}\label{eq:mass},
\end{equation}
where $M$ is the black hole mass, and $\mathcal{M}$ is a constant adopted from the values for the mass function and binary mass ratio in \cite{steeghsetal2013}. For each fit, the randomly selected distance and subsequent inclination and mass were held fixed while the three \middleton{} observations were simultaneously fit with the model \ModelVBhspecSimpl{}. The results from each fit are plotted in Figure \ref{fig:four_panel} which shows the spread in parameter space for mass, distance, and inclination, as well as a histogram of all best-fit $a_*$ obtained. The blue dots are fits which are within 99\% confidence, \chisquared{} $\leq 164.7/125$, and the red dots are fits with \chisquared{} $> 164.7/125$ which highlight regions where fits either became pegged at the maximum spin or the maximum luminosity limit of \bhspec{}. The pile-up of fits at high spin have pegged at the maximum spin limit of \bhspec{} ($a_* = 0.99$), and fits below $a_* \sim 0.5$ signify observation \midc{} has pegged at the luminosity limit of \bhspec{} ($L/L_{\mathrm{Edd}} = 1.77$). The \reid{} preferred values are marked as black X's at our best-fit spin, $a_* = 0.86$, for the \middleton{} observations. The previously assumed values for the mass, distance, and inclination from \middleton{} are marked as green X's at the best-fit spin reported in \middleton{}, $a_* \sim 0.72$. The bottom x-axis is the logarithm of ($1 - a_*$) to better show the portion of moderate to maximal spins, while the top x-axis is just $a_*$. Calculating a 1-sigma confidence interval on either side of our best-fit spin for only the acceptable fits (in blue) gives a spread of $a_* \sim 0.60 - 0.97$.

\begin{figure*}
    \centering
    \includegraphics[width=\textwidth]{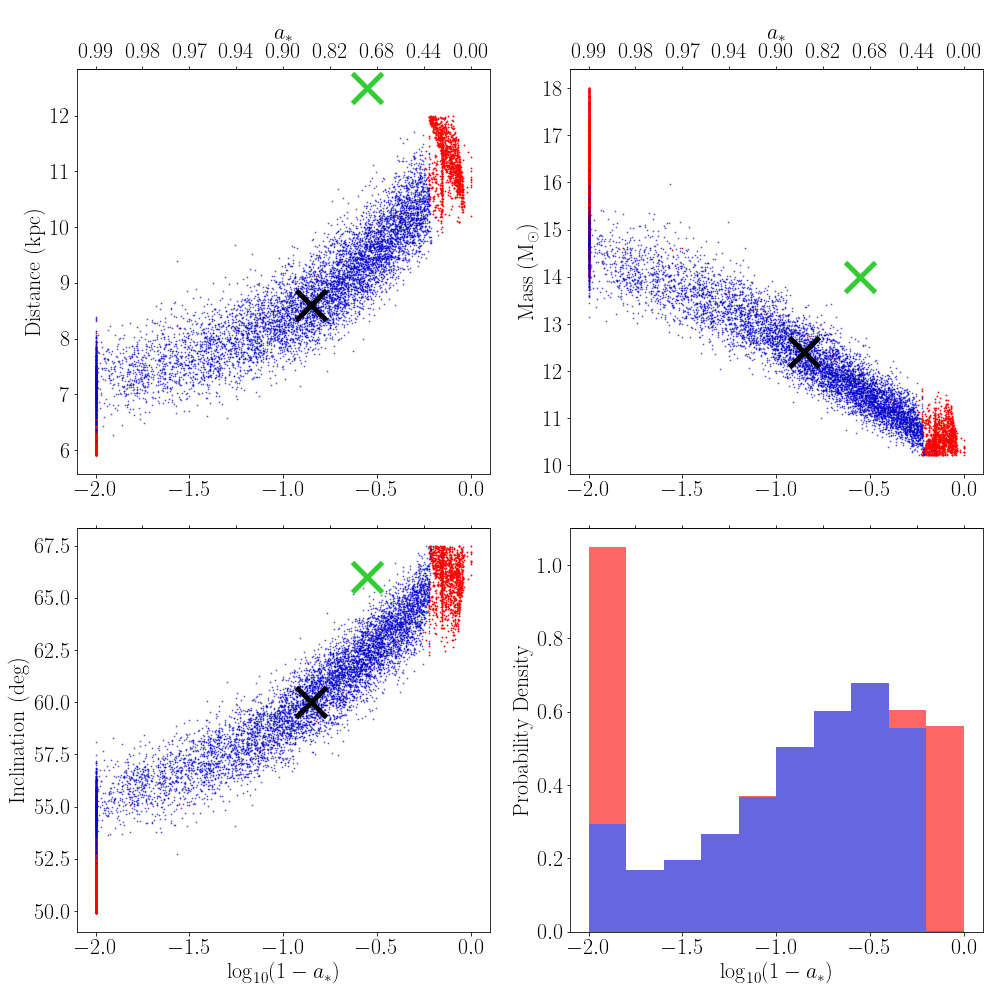}
    \caption{Range of parameter space for the mass, distance, inclination, and resulting best-fit spin, $a_*$, for fits to the three \middleton{} observations, \mida{}, \midb{}, \midc{} tied together and fit with the model \ModelVBhspecSimpl{} (see Section \ref{ss: relativistic}). These fits account for the parameter dependence of inclination and mass on the distance to \GRS{} via equations (\ref{eq:inclination}) and (\ref{eq:mass}) (\reid{}). Each distance was randomly sampled from a Gaussian distribution centered on the \reid{} preferred value $D = 8.6$ kpc. The blue dots indicate fits with \chisquared{} $\leq 164.7/125$ which are within 99\% confidence. The red dots in each panel indicate fits with \chisquared{} $> 164.7/125$ which are outside the 99\% confidence. The pile-up of red dots at high spins is due to fits which have pegged at the maximum spin limit of \bhspec{} ($a_* = 0.99$). The red fits at lower spins $a_* \lesssim 0.5$ indicate fits in which observation \midc{} has pegged at the luminosity limit of \bhspec{} ($L/L_{\mathrm{Edd}} = 1.77$, where $L_{\mathrm{Edd}}$ is the Eddington luminosity). The \reid{} preferred values ($D = 8.6$ kpc, $M = 12.4$ \solarmasses{}, $i = 60$\degree) are marked with a black ``X'' at the best-fit spin for these assumed values, $a_* = 0.863^{+0.014}_{-0.015}$. The best-fit parameter values are listed in Table \ref{tab:bestfit_middleton} for the \reid{} preferred values). The distance, mass, and inclination previously assumed by \middleton{} (12.5 kpc, 14 \solarmasses{}, 66\degree, respectively) are marked with a green ``X'' for comparison. The bottom right panel shows a histogram of all the resulting best-fit spins obtained for the \middleton{} observations.  Note that the histogram is stacked rather than superimposed, and the area under the histogram integrates to 1.}
\end{figure*} \label{fig:four_panel}

The same random sampling continuum-fitting analysis was done for the three \mcclintock{} observations, the results of which are shown in Figure \ref{fig:mcclintock_spin_colormap}. This plot shows the range of distances, masses, and inclinations which produce fits to this dataset within 99\% confidence (\chisquared{} $< 172.66/132$) using the model \ModelPBhspecSimpl{} and the same parameter prescription in Section \ref{ss: relativistic}. Each dot is one realization for a fixed mass, distance, and inclination. When compared to the \middleton{} results from Figure \ref{fig:four_panel}, the \mcclintock{} results show poorer fits for distances below $\sim 10$ kpc, with most of the best-fit spins pegging at the maximum spin limit of \bhspec{} ($a_* \sim 0.99$). There are also a number of poor fits that prefer more moderate spins ($a_* \gtrsim 0.90$), but the \simpl{} scattering fractions become high in these cases and are no longer consistent with disk-dominated results. Note that the range of the x-axes differ from Figure \ref{fig:four_panel} in that the \mcclintock{} fits do not reach spins below $a_* \sim 0.9$.

\begin{figure*}
    \centering
    \includegraphics[width=\textwidth]{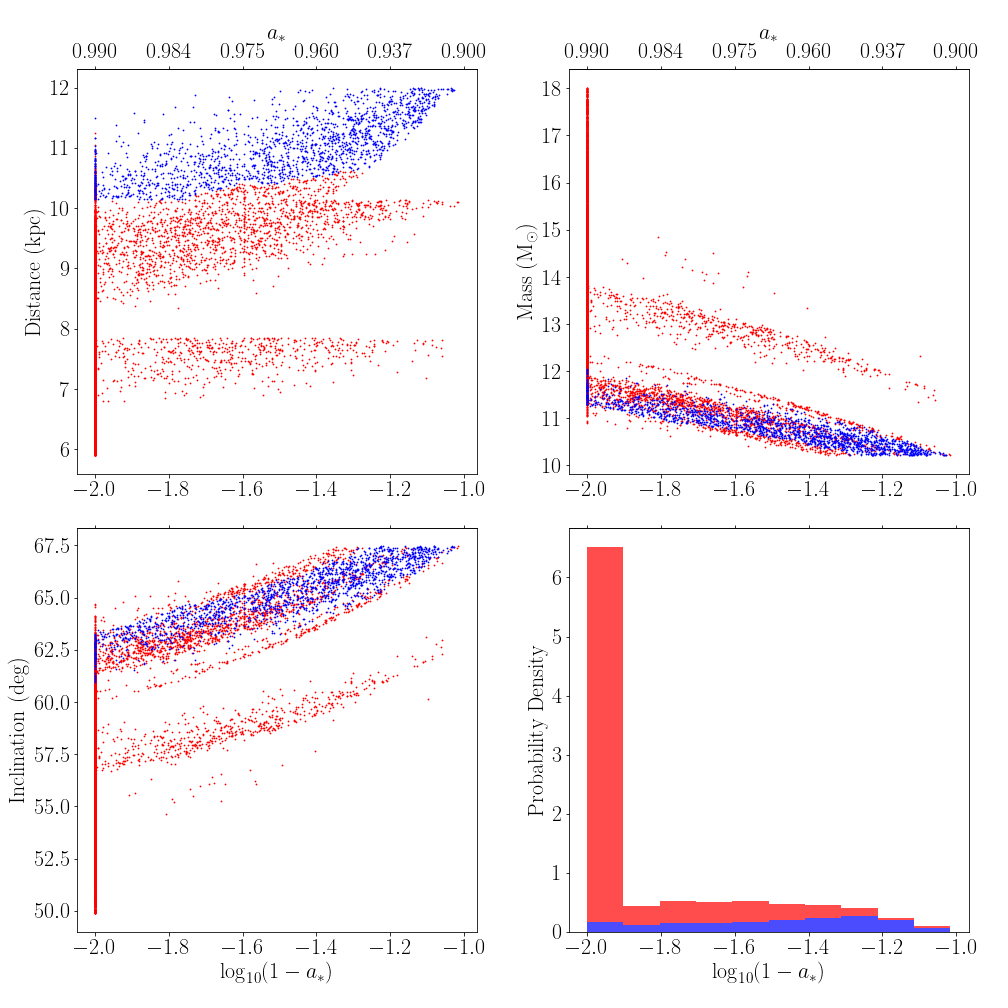}
    \caption{The same analysis in Figure \ref{fig:four_panel} is performed for the three \mcclintock{} observations \mca{}, \mcb{}, \mcc{}. Note that when comparing this figure to Figure \ref{fig:four_panel}, the spin axis here is truncated at $a_* = 0.900$ as there were no best-fit, moderate spin values below this for the \mcclintock{} fits. These observations were similarly fit together with spins tied in the model \ModelPBhspecSimpl{}. Each dot represents one realization for a fixed mass, distance, and inclination. All blue fits shown here are within 99\% confidence (\chisquared{} $\leq 172.7/132$). All red fits have a \chisquared{} $> 172.7/132$ where the \simpl{} scattering fractions have become high and are no longer consistent with disk-dominated results. A pile-up of spins is shown at $a_* = 0.99$ where fits have pegged at the maximum spin limit of \bhspec{}, similar to Figure \ref{fig:four_panel}. The lower right panel shows a stacked histogram of all best-fit spins where the sum over all bins equates to 1.}
\end{figure*} \label{fig:mcclintock_spin_colormap}

\section{Discussion}
\label{s:discuss}

\subsection{Implications for the GRS 1915+105 system}

The relative merits of the \mcclintock{} and \middleton{} data selection were debated in those papers and are summarized in Section~\ref{s:intro}.  We will not discuss this at further length here but we note that one of the objections to the \middleton{} datasets is that their larger Eddington ratios imply thicker accretion disks, potentially invalidating the assumptions of the thin disk model underlying the \bhspec{} model.  With the revised system parameters, the implied Eddington ratios are now slightly lower, with the lowest Eddington observation being in a range where the disk model remains relatively thin. More generally, the relatively high Eddington ratio of \GRS{} is sometimes hypothesized to account for its relatively unique variability, but our best-fit constraints imply the source is generally sub-Eddington or at most slightly super-Eddington.

Although more moderate spins are allowed by the \middleton{} datasets, relatively high spin is implied for the \reid{} best-fit system parameters. Black holes in X-ray binaries are expected to be born with low to moderate spins ($a_* \lesssim 0.7$), although this is subject to uncertainties in the core-collapse process \citep{gammieetal2004}.  It is also not clear that they can be significantly spun up by accretion under standard assumptions about mass transfer \citep{king_kolb1999}.  The high spins inferred here would then imply that either black holes are born with higher natal spin than expected or experience phases of high mass transfer to spin them up.  

It is perhaps notable that the best-fit values from \middleton{} are in the ballpark where general relativistic magnetohydrodynamic simulations suggest magnetic torques would balance the spin-up due to accretion \citep{gammieetal2004}.  This limit ($a_* \lesssim 0.94$) is more stringent than the commonly cited limit of $a_*=0.998$ from \cite{thorne1974}, which only accounts for the angular momentum carried by the radiated photons.  The \middleton{} results are thus consistent with \GRS{} being spun up by accretion and reaching an equilibrium with magnetohydrodynamic torques provided by field lines connected to the black hole, while \mcclintock{} results exceed this nominal limit. Note, however, that the presence of such torques may have an effect on the accretion disk emission \citep{gammie1999,agol_krolik2000,kulkarnietal2011,schnittmanetal2016} and are not accounted for in the present analysis.

Our results for the \middleton{} data are inconsistent with those of \reid{}, who report best-fit $a_* \simeq 0.98$. Accounting for systematic uncertainties, they report $a_* > 0.92$, which is consistent with spin estimates in Figure~\ref{fig:mcclintock_spin_colormap}.  Our results for the \mcclintock{} data are in better agreement in that both analyses favor near maximal spin but differ in that \reid{} managed to find suitable fits for the revised distance of $8.6$ kpc.  This may owe in part to \reid{} reanalyzing a large sample of \RXTE{} observations, selecting all observations that obey a criterion $L/L_{\rm Edd} \le 0.3$, $\chi^2_\nu \le 2$, and $f_{\rm sc} < 0.25$.  It is possible that the \mcclintock{} selected datasets may have been selected out in the process using the revised system parameters.  \citet{sreeharietal2020} also find a near maximal best-fit spin for \kerrbb{} fits to \textit{AstroSat} observations of \GRS{}. These results are notable in that, like \middleton{}, they are for observations that would nominally place the emission above the Eddington limit.  Their analysis differs in allowing the mass to be a free parameter, although their best-fit mass is consistent with the \reid{} constraints.

In addition to the continuum-fitting method, the spin of \GRS{} has also been estimated by fitting the relativistically broadened reflection spectrum due to irradiation of the accretion disk by a corona \citep{blumetal2009,milleretal2013} or via modeling of quasiperiodic oscillations \citep[QPOs;][]{toroketal2011,sramkovaetal2015}.  Although a range of results have been reported with both methods, the reflection fitting efforts are both consistent with high spins $a_* \simeq 0.98$ while the QPO model favors somewhat lower spins $a_* \sim 0.7-0.9$.  Based on previous results, this puts the reflection fitting results in good agreement with \mcclintock{} and the QPO estimates in better agreement with \middleton{}.  For our results, the \mcclintock{} fits are still broadly in agreement with near maximal spin as long as \GRS{} lies at the far end of the distance distribution allowed by VLBI parallax.  The \middleton{} data now provide a larger best-fit spin in better agreement with reflection fitting. More moderate spins are still favored albeit with large uncertainty.

\subsection{Impact and Uncertainties in Interstellar Absorption}
\label{ss: absorption}
Typically when fitting the continuum of a BHXRB, a model for the photoelectric absorption along the line of sight (e.g. \phabs{}, \varabs{}) is needed.  Using the model \varabs{} and abundances from \cite{anders_ebihara1982}, \middleton{} assumed a column value of $n_{\rm{H}} = 4.7 \times 10^{22}$ $\rm{cm}^{-2}$ for all elements except Si and Fe which were fixed at $16.4 \times 10^{22}$ $\mathrm{cm^{-2}}$ and $10.9 \times 10^{22}$ $\mathrm{cm^{-2}}$, respectively.  These values were reported by \cite{leeetal2002} for \textit{Chandra} X-ray observations of \GRS{} assuming ISM abundances.  Relativistic disk reflection studies constraining the spin in \GRS{} report best-fit values which also favor a high absorption column using the \phabs{} model ($n_{\rm{H}}=4.15-5.64 \times 10^{22}$ $\rm{cm}^{-2}$, \citealt{blumetal2009}; $n_{\rm{H}}=6.1 \times 10^{22}$ $\rm{cm}^{-2}$, \citealt{milleretal2013}). \cite{leeetal2002} also reported a S- and Mg-derived hydrogen column assuming solar abundances, giving a more moderate column value of $n_{\rm{H}} \sim 3.1 \times 10^{22}$ $\rm{cm}^{-2}$. This is in better agreement with other modest column estimates from ASCA X-ray observations \citep[$n_{\rm{H}} = 3.8 \times 10^{22}$ $\rm{cm}^{-2}$,][]{ebisawaetal1994} along with millimeter and radio observations \citep[$n_{\rm{H}} = 3.6 \times 10^{22}$ $\rm{cm}^{-2}$,][]{chapuis_corbel2004}. Following these modest estimates, \mcclintock{} adopted a value of $n_{\rm{H}} = 4.0 \times 10^{22}$ $\rm{cm}^{-2}$ (assuming abundances from \citealt{anders_grevesse1989}).

Not only did \mcclintock{} and \middleton{} assume different values for the column, they also chose different XSPEC absorption models. \mcclintock{} selected \phabs{} for their analysis, while \middleton{} chose \varabs{}. We found that when using the same $n_{\rm{H}}$ value and all other variables kept the same, \varabs{} tended toward a lower $a_*$ than \phabs{} did. Fits to the \mcclintock{} data with either \phabs{} or \varabs{} tended to fit better with low $n_{\rm{H}}$ values, while fits to the \middleton{} data tended to fit better with higher $n_{\rm{H}}$ values. An overall trend between both \varabs{} and \phabs{} is that the value for $a_*$ decreased as the $n_{\rm{H}}$ column increased for both datasets. To maintain consistency in our re-analysis we kept the same XSPEC absorption models and values for $n_{\rm{H}}$ chosen by each group, but note the sensitivity and dependence of the spin on the assumed $n_{\rm{H}}$ column value as a source of uncertainty in the $a_*$ estimate in this work.

Aside from the line-of-sight hydrogen column estimates, kinematic studies near \GRS{} have also located a molecular cloud at a distance of $9.4 \pm 0.2$ kpc \citep{chatyetal1996, chapuis_corbel2004}. The new VLBI constraints may have implications for the history and conception of the \GRS{} system given the distance of $8.6$ kpc, which could place it with the observed interstellar structure (\reid{}).

\subsection{Uncertainties in Models and System Parameters}

The \reid{} VLBI parallax measurements provide much stronger constraints on the system parameters than were previously available, but our analysis shows that remaining uncertainties still allow for a rather large range of spins.  If we treat the models as robust to systematic uncertainties, then our fitting constraints nominally imply strict limits on the system parameters. Figure~\ref{fig:four_panel} implies that for relatively low distances and inclinations, the implied spin from the \middleton{} data would be higher than $a_*=0.99$, challenging the theoretical understanding on black hole spin limits.  The constraints are even stronger for the \mcclintock{} data, which would limit the distances to $D \gtrsim 10$ kpc, near the outer limits of what is allowed by the \reid{} constraints.  In fact, this result is consistent with predictions made in Figure~18 of \mcclintock{}, which predicted \GRS{} to lie within an error triangle whose minimum distance was just under $\sim 10$ kpc.

The model for system parameters implied by equations~(\ref{eq:inclination}) and (\ref{eq:mass}) is also subject to systematic uncertainties. First, equation~(\ref{eq:inclination}) assumes the observed superluminal motion can be interpreted as emission from plasma following ballistic trajectories and that these trajectories lie along the spin axis of the black hole.  Furthermore, it is conceivable that the plane of the binary is not perpendicular to the black hole spin axis \citep{fragileetal2001,maccarone2002}, although there are theoretical arguments that such misalignments should typically be modest \citep{fragos_mcclintock2015}.  If such misalignment is present, it seems likely that the inner accretion disk would align with the spin axis due to the action of Lense-Thirring precession \citep{bardeenpetterson1975}.  In this case, our use of the jet to fix the disk inclination would still be reasonable as long as the jet is aligned with the spin axis, although GRMHD simulations of misaligned disks indicate this may not be guaranteed \citep{liskaetal2018} and observations of V404 Cygni show the jet angle to precess \citep{millerjonesetal2019} perhaps due to Lense-Thirring precession associated with the high mass accretion rates inflating the disk \citep{middletonetal2018, middletonetal2019}. The inclination implied by observations of the jet would then not correspond to the binary inclination in equation~(\ref{eq:mass}), and the inferred black hole mass would be incorrect.

An independent constraint on the inclination comes from the reflection spectral modeling, where the relativistic line profiles are sensitive to the viewing inclination.  The best-fit inclinations from \citet{blumetal2009} are $i=55^\circ$ or $i=69^\circ$ depending on the reflection model used. \citet{milleretal2013} found inclinations ranging from $65^\circ$ to $74^\circ$ depending on the model. \citet{blumetal2009} constrained the inclination to lie between $55^\circ$ and $75^\circ$, while \citet{milleretal2013} constrained it to be between $65^\circ$ and $80^\circ$, both based on interpretations of constraints from the superluminal jet model \citep{fenderetal1999}. The allowed inclination ranges are generally higher than those used in our analysis because these papers predate the \reid{} measurements. The higher inclinations ($i \gtrsim 70^\circ$) would pose a challenge to the super-luminal motion interpretation for the new parallax distance, but if we assume they imply that the inclination should be towards the high end of the allowed range, they would push the \middleton{} results to low spins that would be at odds with the best-fit spins from these reflection models.  The \mcclintock{} results could remain consistent with near maximal spin, but again this requires \GRS{} to be located at the more distant end of the range allowed by VLBI parallax measurements.

We emphasize that these constraints are subject to unquantified systematic uncertainties in the underlying accretion disk model.  This could arise from inaccuracies in the underlying thin disk model \citep{shakura_sunyaev1973,novikov_thorne1973} or because of errors in the TLUSTY atmosphere models \citep{davis_hubeny2006}.  The former is perhaps most worrisome for the highest Eddington ratio observations from \middleton{}, where the thin disk assumption would begin to break down.  It is also possible that some aspect of the model is inaccurate, such as the stress prescription \citep{donedavis2008}, assumption of a torque-free inner boundary  \citep{gammie1999,agol_krolik2000} and truncation of emission at the ISCO \citep{krolik_hawley2002,abramowiczetal2010}. The relativistic thin disk models employed here are broadly consistent with GRMHD models, but emission interior to the ISCO and magnetic torques would provide a modest bias towards higher spins when using standard models \citep{kulkarnietal2011,schnittmanetal2016}. In that case, the near maximal best-fit spins might still be indicative of high spin, but not necessarily maximal spin. These models also assume a razor thin disk even though they can be at accretion rates where the thin disk assumption breaks down.  \citet{zhouetal2020} found that considering a model with a finite disk thickness led to a modestly higher spin for fits to \RXTE{} observations of \GRS{}, but their thin disk fits were already near maximal.

The spectra derived from atmosphere modelling with TLUSTY are another potential source of error.  Errors in the atmosphere models and spectra could arise from inaccurate assumptions about the vertical distribution of dissipation, contributions from magnetic pressure support, inhomogeneities in the turbulent disk, and lack of irradiation of the surface \citep{davisetal2005,davisetal2006,davisetal2009,tao_blaes2013,narayanetal2016}. Figure~\ref{fig:color_correction} provides a sense of the degree to which spectral hardening errors would impact our spin results.  Further discussion of the spectral hardening implied by TLUSTY calculations can be found in \citet{davis_el-abd2019} while \citet{salvesen_miller2021} provide a thorough review of uncertainties and quantitative estimates of their impact on spin measurements.

We note that the selection criteria is also a clear source of uncertainty, since two different methods provide nominally disk-dominated spectra that yield different results.  This concern for \GRS{} contrasts with other sources (mainly soft X-ray transient low mass X-ray binaries) that tend to approximately follow a luminosity proportional to temperature to the fourth power relation \citep{gierlinski_done2004,dunnetal2011}.  Since color corrections tend to vary relatively weakly \citep{davis_el-abd2019}, this means different observations of the same source likely yield consistent inner disk radii and consistent spins. \GRS{} tends to be highly variable and less consistent, which is partly why \cite{dunnetal2011} exclude it from their analysis.  Nor is it clear that its variability properties are consistent with other sources in its nominally disk-dominated states, possibly indicating contamination from a warm Comptonizing component \citep{zdziarskietal2005, uedaetal2010}. The quality of fit for our spectral modelling is sensitive to the chosen hard X-ray model. Fits with the \nthcomp{} model can provide a lower \chisquared{} than with the \simpl{} model, and did not provide consistently disk-dominated fits when paired with \bhspec{}. Although the \simpl{} model provides a poorer fit, it provides relatively disk-dominated results, consistent with Comptonization models where the seed photons are provided by the accretion disk and scattered in a hot corona. Further work is needed to robustly characterize the disk-dominated states of \GRS{} and ascertain how this may affect the best-fit spin constraints.

\section{Summary and Conclusions}
\label{s:conclusions}

We re-examine the continuum-fitting based spin estimates for \GRS{} in light of new constraints on the mass, distance, and inclination from VLBI parallax (\reid{}). We find that the discrepancies between data selected by \middleton{} and \mcclintock{} persist, implying that the selection criteria of one (or both) is inconsistent with the assumptions of the thin disk model.  \mcclintock{} showed a trend towards lower spin as the luminosity of the observations increased, indicating the discrepancy may be driven primarily by different Eddington ratio ranges of the two datasets.  The revised system parameters lower the mass, but lead to relatively smaller implied luminosities, leading to lower overall Eddington ratios for fits to both datasets.  This somewhat mitigates concerns that the Eddington ratios in the \middleton{} models were too high, but the highest Eddington ratio is still close to unity ($L/L_{\rm Edd}=0.87$), where the scale height of the disk is unlikely to remain small compared to the radius, as assumed in the model.

The new system parameters drive both datasets to higher spins. Since \mcclintock{} were already fitting for near maximal spin, this presents a challenge. For the \bhspec{} model (or \kerrbb{} model with color corrections set to match \bhspec{}), we cannot obtain a good fit for the preferred (\reid{}) system parameters.  Good fits to these data can only be found if the color correction is allowed to vary to values significantly higher ($f \gtrsim 2.2$) than implied by \bhspec{} or the distance to \GRS{} is near or greater than 10 kpc, consistent with the prediction of \mcclintock{} (their Figure~18).  The spin would remain near maximal ($a_* \simeq 0.99$) for a distance of 10 kpc, consistent with constraints from modeling of the reflection spectrum.

For the \middleton{} data, the best-fit spin is moderately high ($a_* \simeq 0.86$) for the best-fit \reid{} system parameters.  We find, however, that a fairly broad range of spins are allowed when the uncertainty in the parallax distance and jet model inclination are accounted for, as indicated by Figure~\ref{fig:four_panel}.  In principle, this allows the spin to match constraints from either the near maximal spins from reflection modeling or the more moderate spins from QPO models.  Near maximal spin, however, would result for fairly low inclinations in our model, which would be inconsistent with the best-fit inclinations from reflection modelling.  The low end of the allowed spin distribution is sensitive to the maximum Eddington ratio permitted by \bhspec{}  ($L/L_{\rm Edd} = 1.77$).  Therefore, the lower limit on $a_{*}$ is tied to the Eddington ratio beyond which one says the thin disk model is no longer valid.

Although the VLBI parallax measurements are an impressive achievement, our results indicate that even stronger constraints are necessary to provide a tight constraint on the spin with the continuum-fitting method, and to help resolve the discrepancies driven by data selection.  We note that such constraints also have implications for the reflection spectrum modeling through the dependence of relativistic Doppler shift and beaming on the observer inclination.  This analysis would also benefit from better constraints on the interstellar absorption toward \GRS{}.  The datasets modeled here prefer different models for the absorption and are particularly sensitive to the hydrogen column assumed.  The range of columns used in the literature vary by more than a factor of two, which is enough to modify the best-fit spin, with larger assumed columns generally providing lower spins.

\section*{Acknowledgements} We thank the anonymous reviewer for their helpful suggestions and making this paper clear and more consistent. We thank Tom Maccarone, Evan Smith, and Jack Steiner for useful discussions and helpful comments. This work also benefited from SWD's many discussions with Jeff McClintock about the application of the continuum-fitting method and we wish he was still with us to provide input on this work and continue the efforts himself. This work was supported by NASA Astrophysics Theory Program grant 80NSSC18K1018 and assisted by the Jefferson Scholars Foundation Fellowship.

\bibliography{bibliography.bib}

\end{document}